\documentclass[twocolumn,showpacs,preprintnumbers,amsmath,amssymb,prl,longbibliography,superscriptaddress]{revtex4-1}
\usepackage{bbm}
\usepackage{mathrsfs}
\usepackage{graphicx}
\usepackage{dcolumn}
\usepackage{bm}
\usepackage{amsmath}
\usepackage{amsfonts}
\usepackage{color}
\usepackage[colorlinks=true,citecolor=blue,anchorcolor=blue]{hyperref}
\usepackage{floatrow}
\begin{document}

\title{Flat Chern Band From Twisted Bilayer MnBi$_2$Te$_4$}
\author{Biao Lian}
\thanks{biao@princeton.edu}
\affiliation{Princeton Center for Theoretical Science, Princeton University, Princeton, New Jersey 08544, USA}
\author{Zhaochen Liu}
\affiliation{State Key Laboratory of Surface Physics, Department of Physics, Fudan University, Shanghai 200433, China}
\author{Yuanbo Zhang}
\affiliation{State Key Laboratory of Surface Physics, Department of Physics, Fudan University, Shanghai 200433, China}
\affiliation{Institute for Nanoelectronic Devices and Quantum Computing, Fudan University, Shanghai 200433, China}
\author{Jing Wang}
\thanks{wjingphys@fudan.edu.cn}
\affiliation{State Key Laboratory of Surface Physics, Department of Physics, Fudan University, Shanghai 200433, China}
\affiliation{Institute for Nanoelectronic Devices and Quantum Computing, Fudan University, Shanghai 200433, China}

\begin{abstract}
We construct a continuum model for the Moir\'e superlattice of twisted bilayer MnBi$_2$Te$_4$, and study the band structure of the bilayer in both ferromagnetic (FM) and antiferromagnetic (AFM) phases. We find the system exhibits highly tunable Chern bands with Chern number up to $3$. We show that a twist angle of $1^\circ$ turns the highest valence band into a flat band with Chern number $\pm1$ that is isolated from all other bands in both FM and AFM phases. This result provides a promising platform for realizing time-reversal breaking correlated topological phases, such as fractional Chern insulator and $p+ip$ topological superconductor. In addition, our calculation indicates that the twisted stacking facilitates the emergence of quantum anomalous Hall effect in MnBi$_2$Te$_4$.
\end{abstract}

\date{\today}

\maketitle

Topology has become one of the central topics in condensed matter physics.
The discovery of topological insulator (TI)~\cite{kane2005a,kane2005b,bernevig2006d,koenig2007,fu2007,Chen2009,zhang2009,xia2009,hasan2010,qi2011,wang2017c}, quantum anomalous Hall (QAH) effect~\cite{haldane1988,liu2008,yu2010,chang2013b,wang2013a,wang2015d,liu2016} and other topological states have significantly enriched the variety of quantum matter, and may lead to potential applications in electronics and quantum computation~\cite{ivanov2001,kitaev2003,nayak2008,alicea2011,lian2018b}. Electron-electron interaction plays an essential role in fractional quantum Hall effect, and there have been proposals of strongly correlated topological states such as fractional TI and fractional Chern insulator (FCI) without magnetic field~\cite{levin2009,maciejko2010,qi2011f,tang2011,sun2011,neupert2011,stern2016,spanton2018}. Experimentally realizing such states is, however, challenging because flat topological electronic bands are generally required for electron-electron interactions to manifest.

Recently, it is shown that Moir\'e superlattices in twisted or lattice mismatched two-dimensional (2D) materials can give rise to flat topological bands. A prime example is twisted bilayer graphene (tBLG)~\cite{bistritzer2011,cao2018,yankowitz2018,lu2019}, where the lowest two bands carry a fragile topology~\cite{songz2018,po2018b,ahn2018,lian2019b,xief2019} and become flat near the magic twist angle $\theta\approx1.1^\circ$. In addition, flat valley Chern bands can be realized in tBLG with aligned hBN substrate~\cite{bultinck2019,sharpe2019,serlin2019}, twisted double bilayer graphene~\cite{liux2019,cao2019a,shen2019}, ABC trilayer graphene on hBN~\cite{zhangy2018,chengr2019,chengr2019b} and twisted bilayer transition metal dichalcogenides~\cite{wu2019,jin2019}, etc. The small bandwidths make electron-electron interactions important~\cite{kang2018b,koshino2018,po2018,dodaro2018,xiey2019,kerelsky2019,choiy2019,jiang2019}, and further lead to intriguing interacting phases in experiments including superconductivity, correlated insulator and QAH effect.

So far, all of the experimental Moir\'e systems are time-reversal (TR) invariant at the single particle level, thus the total Chern number always equals to zero. Therefore, even with flat bands, it is difficult to achieve TR breaking interacting topological states such as the FCI in these systems. This motivates us to consider the Moir\'e superlattice of TR breaking layered materials. A promising system is 3D antiferromagnetic (AFM) topological axion insulator MnBi$_2$Te$_4$~\cite{zhangd2019,lij2019,gong2019,otrokov2018,otrokov2019,leeh2018,yan2019,chenb2019,deng2019,liuc2019,ge2019,hao2019,chen2019,li2019,swatek2019,li2010,wang2011,wang2016a,rani2019}, which can be driven into a ferromagnetic (FM) Weyl semimetal or 3D QAH insulator. The material consists of Van der Waals coupled septuple layers (SLs) and is FM within each SL. Few-SL MnBi$_2$Te$_4$ films have been shown to host instrinsic QAH effects~\cite{deng2019}.

In this letter, we study the band structure of twisted bi-SL MnBi$_2$Te$_4$ (tBMBT) Moir\'e superlattice as an example of TR breaking Moir\'e systems. The magnetization of the two SLs may be either the same (FM) or opposite (AFM), both of which are explored here. We find the band structure contains a number of nondegenerate Chern bands, which undergo Chern number topological phase transitions with respect to tunable system parameters such as the twist angle, staggered layer potential and the magnetization. In particular, by tuning staggered layer potential, one can drive the first valence band of both FM and AFM tBMBT into a flat Chern band with Chern number $\pm1$ around twist angle $1^\circ$, which is energetically separated from the other bands. tBMBT thus provides an ideal platform for searching for FCI and other TR breaking interacting topological phases. In addition, low energy bands with Chern number higher than $\pm1$ may also be realized by tuning the parameters.

\begin{figure}[t]
\begin{center}
\includegraphics[width=3.4in,clip=true]{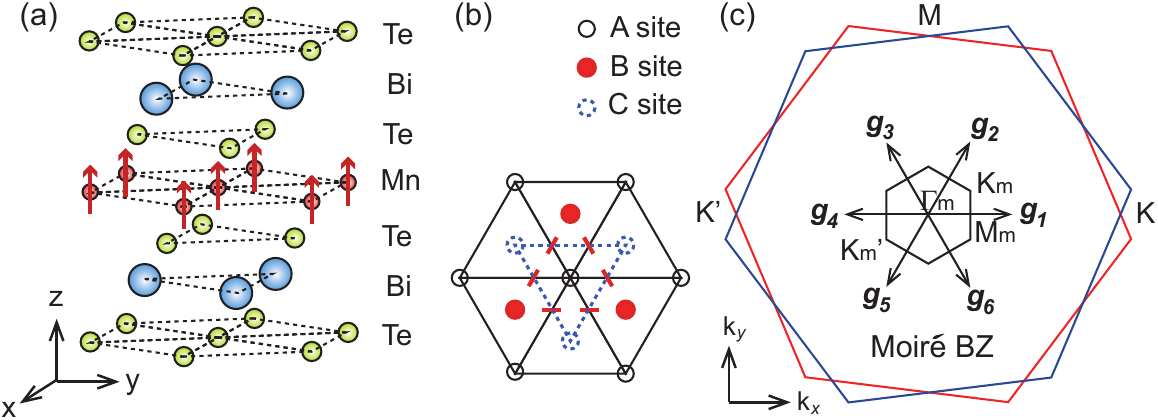}
\end{center}
\caption{(a) A single SL of MnBi$_2$Te$_4$, where seven atomic layers form the ABC-type stacking. (b) Top view illustration of A, B, and C stacking configurations of the triangular atomic lattices. (c) The relatively twisted single SL BZ and the Moir\'e BZ of tBMBT.
}
\label{figMoire}
\end{figure}

The bulk MnBi$_2$Te$_4$ has a layered rhombohedral crystal structure with the space group $D_{3\text{d}}^5$ (No.~166). Each unit cell consist of seven-atom layers (Te-Bi-Te-Mn-Te-Bi-Te) arranged along the trigonal $z$-axis with the ABC-type stacking, referred to as an SL, as shown in Fig.~\ref{figMoire}(a). The in-plane triangular lattice constant is $a_0=4.334$~\AA, and the thickness of a unit cell (consisting of 3 SLs) is $c_0=40.91$~\AA. Neighboring SLs have van der Waals couplings, and the adjacent atomic layers of neighboring SLs form AB stacking in the ground state crystal structure.


Below a N\'{e}el temperature of $\sim25$~K, each SL of the bulk MnBi$_2$Te$_4$ develops an intralayer FM order on the Mn atoms with an out-of-plane easy axis, but adjacent SLs couple anti-parallel to each other, yielding a topological axion insulator with an out-of-plane layered AFM order. 
The FM phase with an out-of-plane easy axis is a competing ground state with a slightly higher energy, where the system is a Weyl semimetal or a 3D QAH insulator~\cite{lij2019,zhangd2019}. The intrinsic magnetism and band inversion make it highly promising to realize the intrinsic QAH effect in few-SL MnBi$_2$Te$_4$ thin films~\cite{zhangd2019,lij2019,otrokov2019,deng2019,liuc2019,ge2019}.


The weak Van der Waals coupling between SLs allows the implementation of tBMBT by stacking two mono-SLs with a twist angle. The first-principles calculations show that few-SL MnBi$_2$Te$_4$ have competing FM and AFM ground states~\cite{zhangd2019,lij2019,otrokov2019}. While the AFM phase is more likely, it may be flipped into FM by a $2\sim4$T magnetic field \cite{otrokov2018,deng2019,suppl} or top/bottom FM heterostructure proximities.  
Therefore, we investigate both the FM and AFM phases of tBMBT, where the two SLs have the same and opposite $z$ direction FM orders, respectively.

\emph{Model}. We now construct an effective continuum model~\cite{bistritzer2011} for tBMBT formed by two SLs stacked on top of each other with a twist angle $\theta$, which is generic for $C_{3z}$ symmetric layered magnetic materials with low energy Dirac electrons. The Hamiltonian for such a model can be written in real space as
\begin{equation}\label{Hmoire}
H=
\begin{pmatrix}
h_{1,\frac{\theta}{2}}(-i\nabla)+U_d & T(\mathbf{r})\\
T^\dag(\mathbf{r}) & h_{2,-\frac{\theta}{2}}(-i\nabla)-U_d\\
\end{pmatrix},
\end{equation}
where $-i\nabla$ is the 2D momentum in the monolayer Brillouin zone (BZ) of each SL, $h_{l,\pm\frac{\theta}{2}}$ is the $4\times 4$ monolayer Hamiltonian of the $l$-th SL ($l=1,2$) rotated by angle $\pm\theta/2$, $U_d$ is a staggered layer potential which can be tuned by the top and back gates, and $T(\mathbf{r})$ is the $4\times 4$ interlayer Moir\'e hopping potential. The basis of the monolayer Hamiltonian $h_{l,\pm\frac{\theta}{2}}$ is $(|p_{z,\text{Bi}}^{+},\uparrow\rangle, |p_{z,\text{Te}}^{-},\downarrow\rangle, |p_{z,\text{Te}}^{-},\uparrow\rangle, |p_{z,\text{Bi}}^{+},\downarrow\rangle)^T$ of the $l$-th SL ($l=1,2$), where superscripts ``$+$'', ``$-$'' stand for parity. $|p_{z,\text{Bi}}^{+},s\rangle$ is the spin $s$ bonding state of the $p_z$ orbitals of two Bi layers, and $|p_{z,\text{Te}}^{-},s\rangle$ is the spin $s$ antibonding state of the two $p_z$ orbitals of the top and bottom Te layers. Since the low energy physics in MnBi$_2$Te$_4$ is located near the $\Gamma$ point, we set the origin of the momentum $-i\nabla$ to be $\Gamma$ of the monolayer BZ. In the below, we study the FM and AFM phases separately.

\emph{FM phase}. Depending on the strength of FM exchange field, the untwisted FM bilayer MnBi$_2$Te$_4$ may be either a QAH insulator of Chern number $\pm1$, or a trivial insulator which enters the QAH phase under a small magnetic field \cite{deng2019,liuc2019}. To include both possibilities, we introduce a dimensionless FM strength tuning parameter $\gamma_f$, where we fix $|\gamma_f|=1$ to be the critical FM order strength above (below) which the untwisted FM bilayer MnBi$_2$Te$_4$ is a QAH (trivial) insulator \cite{suppl}. Experimentally, $\gamma_f$ is tunable by the magnetic field.

The monolayer Hamiltonian in Eq.~(\ref{Hmoire}) for a FM tBMBT with FM strength $\gamma_f$ can be written as
\begin{equation}\label{htFM}
h_{l,\pm\frac{\theta}{2}}(\mathbf{k})=R^\dag_{\pm\frac{\theta}{2}}\left[h_\text{N}(\mathbf{k})+ \gamma_f h_{\text{FM}}(\mathbf{k})\right]R_{\pm\frac{\theta}{2}},
\end{equation}
where $\mathbf{k}=(k_x,k_y)$ is the 2D electron momentum, $R_{\pm\frac{\theta}{2}}=\text{diag}(e^{\pm i\theta/4},e^{\mp i\theta/4},e^{\mp i\theta/4},e^{\pm i\theta/4})$ is the angle $\pm\theta/2$ rotation matrix about the $z$ axis. $h_\text{N}(\mathbf{k})$ and $h_{\text{FM}}(\mathbf{k})$ are the nonmagnetic part and FM part of the $\mathbf{k}\cdot\mathbf{p}$ Hamiltonian of single SL MnBi$_2$Te$_4$ at the $\Gamma$ point, respectively, which take the forms
\begin{equation}\label{HFmono}
h_\text{N}(\mathbf{k})=\epsilon_0(\mathbf{k})+
\begin{pmatrix}
m(\mathbf{k})& \alpha k_-&&\\
\alpha k_+& -m(\mathbf{k}) &&\\
&&-m(\mathbf{k})&\alpha k_-\\
&&\alpha k_+&m(\mathbf{k})\\
\end{pmatrix},
\end{equation}
and
\begin{equation}\label{FMa}
h_{\text{FM}}(\mathbf{k})=
\begin{pmatrix}
m_{1}(\mathbf{k})& \alpha' k_-&&\\
\alpha'k_+& -m_2(\mathbf{k}) &&\\
&&m_2(\mathbf{k})&-\alpha' k_-\\
&&-\alpha' k_+&-m_1(\mathbf{k})\\
\end{pmatrix}.
\end{equation}
Here $\epsilon_0(\mathbf{k})=\gamma\mathbf{k}^2$ is the particle-hole asymmetry term proportional to the identity matrix, $k_\pm\equiv k_x\pm ik_y$, $m(\mathbf{k})=m_0+\beta_0 \mathbf{k}^2$, and $m_j(\mathbf{k})=m_j+\beta_j\mathbf{k}^2$ ($j=1,2$).

The interlayer Moir\'e hopping potential $T(\mathbf{r})$ is spatially periodic. To the lowest order, it can be Fourier expanded as
\begin{equation}
T(\mathbf{r})=T_0+\sum_{j=1}^6 T_j e^{i\bm{g}_j\cdot\mathbf{r}}\ ,
\end{equation}
where $\bm{g}_j$ ($1\le j\le 6$) are the six smallest Moir\'e reciprocal vectors with length $|\bm{g}_j|=8\pi\sin(\theta/2)/\sqrt{3}a_0$ as shown in Fig.~\ref{figMoire}(c). $\mathbf{r}=0$ is defined as an AA stacking center, where the adjacent atomic layers of two SLs form AA stacking. The matrices can be divided into
\begin{equation}\label{Tj}
T_j=T_j^\text{N}+\gamma_fT_j^\text{FM}, \quad (0\le j\le 6)
\end{equation}
where $T_j^\text{N}$ and $T_j^\text{FM}$ are the nonmagnetic part and FM part, respectively.
The form of matrices $T_j$ and the parameters for the FM phase estimated from bulk calculations are given in the Supplementary Material (SM)~\cite{suppl}.

We now investigate the Moir\'e band structure of the FM tBMBT with respect to $\theta$, $U_d$ and $\gamma_f$.
To distinguish from the original monolayer BZ, we denote the high symmetry points of the hexagonal Moir\'e BZ as $\Gamma_m$, $K_m$ and $M_m$. The bands of FM tBMBT are generically nondegenerate, many of which carry nonzero Chern numbers. The FM tBMBT has $C_{3z}$ and $C_{2x}\mathcal{T}$ symmetries at $U_d=0$ ($\mathcal{T}$ for TR). A nonzero $U_d$ is odd under $C_{2x}\mathcal{T}$ and thus breaks $C_{2x}\mathcal{T}$. Since the Hall conductance $\sigma_{xy}$ is invariant under $C_{2x}\mathcal{T}$, the band Chern numbers of FM tBMBT are invariant under $U_d\rightarrow -U_d$.

\begin{figure}[tbp]
\begin{center}
\includegraphics[width=3.4in,clip=true]{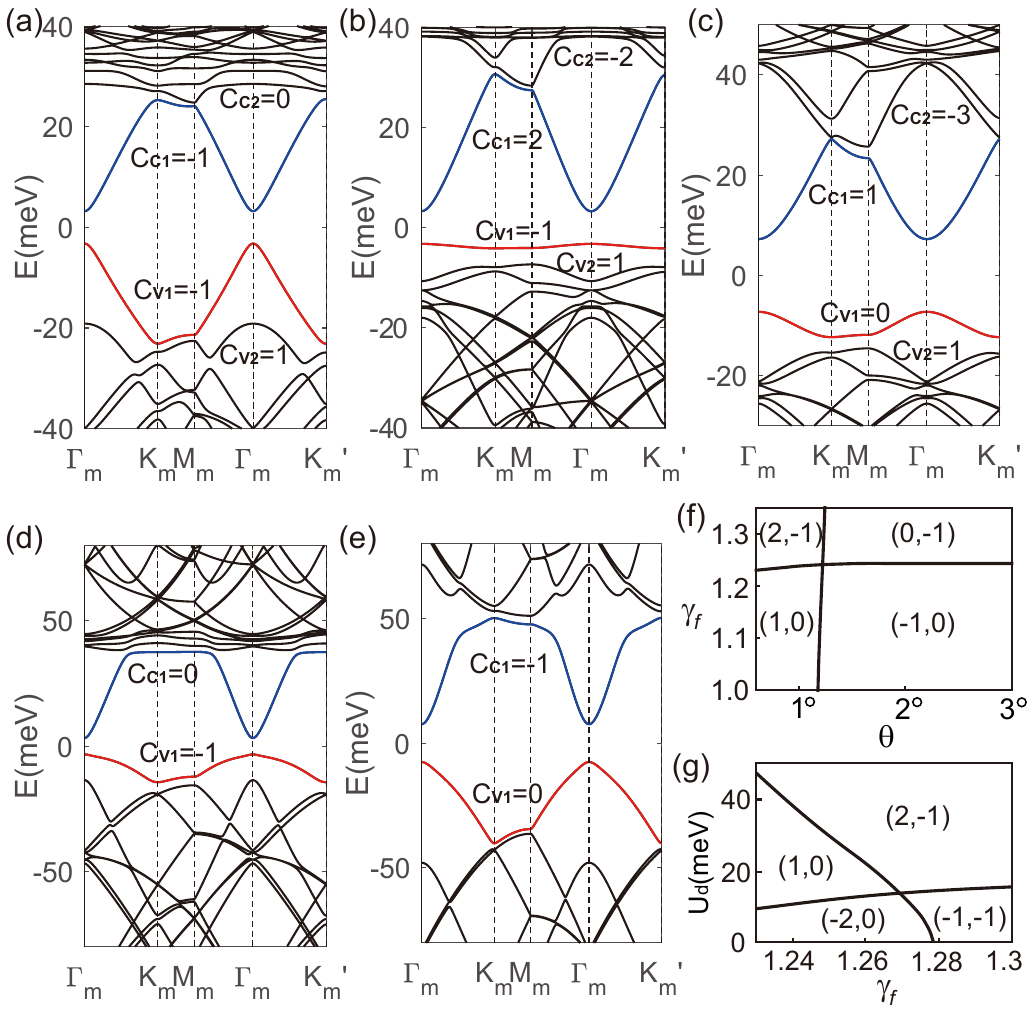}
\end{center}
\caption{The band structure of the FM tBMBT for (a) $\theta=1^\circ$, $U_d=10$~meV, $\gamma_f=1.02$, (b) $\theta=1^\circ$, $U_d=40$~meV, $\gamma_f=1.02$, (c) $\theta=1^\circ$, $U_d=40$~meV, $\gamma_f=0.75$, (d) $\theta=2^\circ$, $U_d=40$~meV, $\gamma_f=1.02$ and (e) $\theta=3^\circ$, $U_d=40$~meV, $\gamma_f=0.75$. 
(f) Chern numbers of the first conduction and valence bands ($C_{C1}, C_{V1}$) as a function of angle $\theta$ and exchange field strength $\gamma_f$, where $U_d=40$~meV is set. (g) ($C_{C1}, C_{V1}$) for $\theta=1^\circ$ as a function of $\gamma_f$ and staggered layer potential $U_d$.}
\label{figFM}
\end{figure}

Fig.~\ref{figFM}(a)-(e) show typical examples of the FM tBMBT Moir\'e band structures, where the Chern number of the $j$-th conduction (valence) band is denoted by $C_{Cj}$ ($C_{Vj}$), and the parameters are given in the caption. The charge neutrality point (CNP) is set as zero. In general, the Chern numbers of the lowest several bands are tunable up to $\pm3$.
However, most bands except for the first conduction and valence bands have no indirect gaps among each other. Therefore, the system is metallic with nonzero Fermi surface Berry phases at high fillings.


Here we mainly focus on the first conduction and valence bands of the FM phase. In the parameter space of $\theta$, $U_d$ and $\gamma_f$, they undergo multiple Chern number topological phase transitions via gap closings at high symmetry points. Fig.~\ref{figFM}(f) shows the Chern number phase diagram of the first conduction and valence bands ($C_{C1},C_{V1}$) with respect to $\theta$ and $\gamma_f$ at fixed $U_d=40$~meV. The gap between the the first conduction and valence bands closes at $\Gamma_m$ point around $\gamma_f=0.93$ for a wide range of $\theta$, which leads to an exchange of Chern number $1$ between these two bands. Accordingly, the FM tBMBT at the CNP is a QAH insulator with Chern number $-1$ when $\gamma_f>0.93$, and the first valence band carries Chern number $C_{V1}=-1$. Therefore, the FM tBMBT enters the QAH phase at a smaller FM strength $\gamma_f$ than the untwisted FM bilayer MnBi$_2$Te$_4$, which suggests that twisting helps achieve the QAH effect in bilayer MnBi$_2$Te$_4$. In addition, the first conduction band undergoes a gap closing with the second conduction band at $K_M$ and $K_M'$ points at angle $\theta\approx 1.2^\circ$ as shown in Fig.~\ref{figFM}(f), where its Chern number changes from $0$ to $2$.

Fig.~\ref{figFM}(g) shows the phase diagram with respect to $U_d$ and $\gamma_f$ at fixed angle $\theta=1^\circ$. As one can see,  adding a staggered layer potential $U_d$ also helps achieve the QAH effect of Chern number $-1$ at the CNP, and accordingly $C_{V1}=-1$. Besides, the Chern number of the first conduction band changes by $3$ at $U_d\approx 10$~meV, which is induced by the gap closing between the first and second conduction bands at three $M_m$ points.

In particular, the first valence band of the FM tBMBT with Chern number either $-1$ or $0$ can be made extremely flat, and the band is energetically separated from other bands near twist angle $\theta=1^\circ$. It is therefore promising to realize TR breaking interacting topological states such as the FCI and the $p+ip$ chiral topological superconductor (TSC). Generally speaking, adding a staggered layer potential $U_d$ flattens the first valence band but not the first conduction band, due to the particle-hole asymmetric term $\epsilon_0(\mathbf{k})$ in Eq.~(\ref{HFmono}). Fig.~\ref{figFM}(a) and (b) show the band structures at $\theta=1^\circ$ and $\gamma_f=1.02$ with $U_d=10$~meV and $40$~meV, respectively, where the first valence band has Chern number $C_{V1}=-1$, and the system has Chern number $-1$ when the Fermi level is at CNP.
In particular, when $U_d=40$~meV in Fig.~\ref{figFM}(b), the bandwidth of the first valence band is suppressed down to $W\approx1$~meV, while its gap with the other nearest bands is $\Delta\approx4$~meV. Such an isolated flat Chern band is therefore an ideal platform for realizing the FCI, where the electron filling is readily tuned by a gate. For an estimation, taking the dielectric constant of the MnBi$_2$Te$_4$ film $\epsilon_r\approx10$, one obtain a Coulomb interaction energy $U\approx6$~meV for filling in the first tBMBT band, which easily exceeds the bandwidth and thus make the FCI possible. Besides, the FM strength $\gamma_f$ can further tune the Chern number of the first valence band and accordingly the Chern number at CNP. Fig.~\ref{figFM}(c) shows the bands at $\theta=1^\circ$, $\gamma_f=0.75$ and $U_d=40$~meV, where both the first valence band and the CNP gap have Chern number $0$. In this case, the first valence band realizes a topologically trivial flat band of bandwidth smaller than $5$~meV.

With either Chern number $0$ or $-1$, the nondegenerate flat valence band allows a single Fermi surface with large density of states when partially filled, leading to a chance of realizing an intrinsic $p+ip$ chiral TSC if a nodeless pairing is developed \cite{read2000,qi2010b,wang2015c,he2017}. The superconductivity experimentally discovered in other Moir\'e systems suggest that superconductivity is more likely to occur in the presence of Moir\'e superlattices~\cite{volovik2018}, where one possible mechanism is the Moir\'e pattern enhances electron-phonon coupling if the superconductivity is phonon induced~\cite{wu2018,lian2018,choi2018}. Therefore, the $p+ip$ TSC might be more achievable in TR breaking Moir\'e superlattices such as tBMBT here than other TR breaking systems.

When $\theta$ is far from $1^\circ$, it is difficult to obtain energetically separated flat bands. For smaller $\theta$, the bandwidths are smaller, but there are hardly indirect gaps except for the CNP gap. For larger $\theta$, not only indirect gaps are rare, but also the bands become more dispersive, as shown in the two examples of Fig.~\ref{figFM}(d) and \ref{figFM}(e) at $\theta=2^\circ$ and $3^\circ$ with $U_d=40$~meV, respectively. Detailed examination reveals that the optimal angles for flat bands in the FM tBMBT fall within $0.8^\circ\lesssim \theta\lesssim 1.2^\circ$.


\emph{AFM phase}. The monolayer Hamiltonian of the $l$-th layer ($l=1,2$) in Eq.~(\ref{Hmoire}) for the AFM tBMBT takes the form
\begin{equation}
h_{l,\pm\frac{\theta}{2}}(\mathbf{k})=R^\dag_{\pm\frac{\theta}{2}}\left[h_\text{N}(\mathbf{k})-(-1)^l \gamma_{af}h_{\text{AFM}}(\mathbf{k})\right] R_{\pm\frac{\theta}{2}},
\end{equation}
where $h_\text{N}(\mathbf{k})$ is still given in Eq.~(\ref{HFmono}) but with different parameters from FM phase, and the AFM term is approximated as
\begin{equation}
h_{\text{AFM}}(\mathbf{k})=\text{diag}(m_1,-m_2,m_2,-m_1)\ ,
\end{equation}
which has no $\mathbf{k}$ dependence. $\gamma_{af}$ tunes the AFM order strength ($\gamma_{af}=1$ represents the strength estimated from the first-principles calculations).
The interlayer Moir\'e potential only contains the nonmagnetic part of Eq. (\ref{Tj}), i.e., $T_j=T_j^N$. The matrices $T_j$ and the parameters for the AFM phase are listed in the SM~\cite{suppl}. In contrast to the 3D AFM MnBi$_2$Te$_4$ which has two-fold degenerate bands protected by the $\mathcal{PT}$ symmetry ($\mathcal{P}$ for inversion), the AFM tBMBT has nondegenerate bands, since the twist angle breaks the $\mathcal{PT}$ symmetry. It only has $C_{3z}$ and $C_{2x}$ symmetries at $U_d=0$, and $C_{2x}$ is further broken when $U_d$ is nonzero.

\begin{figure}[tbp]
\begin{center}
\includegraphics[width=3.4in]{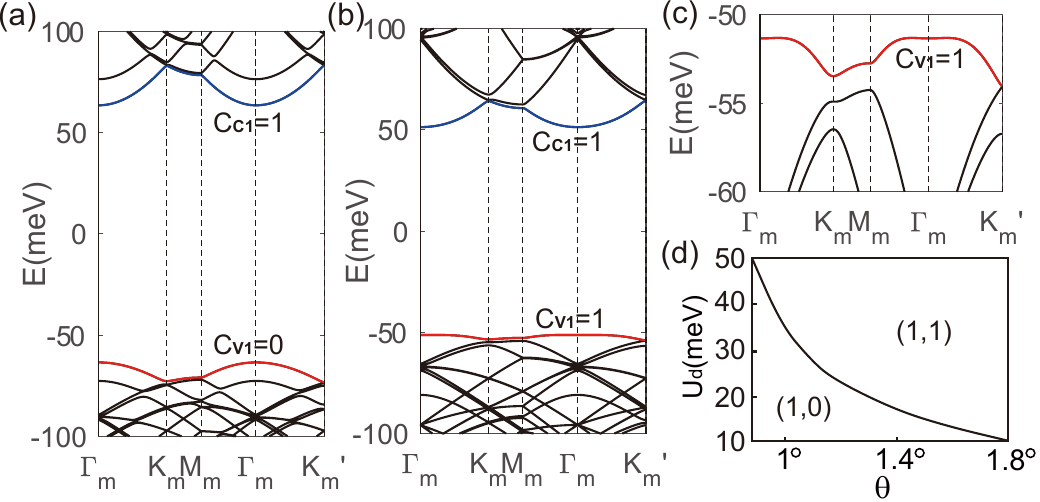}
\end{center}
\caption{The band structure of the AFM tBMBT for (a) $\theta=1^\circ$, $\gamma_{af}=1$, $U_d=10$~meV, and (b) $\theta=1^\circ$, $\gamma_{af}=1$, $U_d=40$~meV.
(c) Zoom-in plot of the valence band structure in (b), showing the bandwidth of the first valence band. (d) ($C_{C1}, C_{V1}$) as a function of the twist angle $\theta$ and the staggered layer potential $U_d$.}
\label{figAFM}
\end{figure}

Since $\sigma_{xy}$ is odd under $C_{2x}$, all the bands of the AFM tBMBT have Chern number zero at $U_d=0$. Nonzero Chern numbers can only arise at nonzero $U_d$, and are odd under $U_d\rightarrow-U_d$. Fig.~\ref{figAFM}(a) and~\ref{figAFM}(b) show the band structure of AFM tBMBT at $\theta=1^\circ$, $\gamma_{af}=1$ for $U_d=10$~meV and $40$~meV, respectively. Similar to the FM phase, increasing $U_d$ flattens the first valence band but not the first conduction band. Besides, the Chern number of the first valence band undergoes a transition from $0$ to $1$ as $U_d$ increases, which is induced by the gap closing at $K_m'$ point (note that $K_m$ and $K_m'$ are not symmetric). Fig.~\ref{figAFM}(d) shows the Chern number phase diagram of the first conduction and valence bands with respect to $\theta$ and $U_d$. Therefore, the first valence band of the AFM tBMBT can also be driven into a flat Chern band separated from the other bands. Fig.~\ref{figAFM}(c) shows a zoom-in plot of Fig.~\ref{figAFM}(b), where the first valence band has a small bandwidth around $W\approx3$~meV, but with a smaller gap to second valence band. The CNP gap always has Chern number $0$. The conclusions are qualitatively insensitive to $\gamma_{af}$. This allows the realization of the QAH effect with Chern number $\pm1$ in the AFM tBMBT by fully emptying the first valence band. More importantly, this indicates it is also possible to realize FCI and other interacting topological phases in the AFM tBMBT. Again, we find the optimal angle for realizing energetically isolated flat bands in AFM tBMBT is around $\theta=1^\circ$. It is worth mentioning that a larger $\theta$ can lead to relatively flat first valence band with $C_{V1}=-2$ but without indirect gap to higher bands~\cite{suppl}.

\emph{Discussion}. The tBMBT with a twist angle near $1^\circ$ host isolated Moir\'e Chern bands, whose bandwidth is significantly smaller than the Coulomb repulsion energy ($2\lesssim U/W\lesssim6$). Mechanically robust single SL of MnBi$_2$Te$_4$ has been obtained experimentally \cite{deng2019}, making it possible to implement tBMBT. The broad variety of tuning parameters including twist angle, staggered layer potential, electron filling, magnetic field, and hydrostatic pressure makes tBMBT a promising platform for realizing the correlated topological phases. The FM phase is more favored than the AFM phase for the flat Chern band to have a larger gap to other bands. Disorders also inevitably exist in realistic materials. Short range scatters will broaden the bandwidth, and thus reduce $U/W$, but the correlated topological phases should be robust against long-range potential fluctuations (i.e. charge puddles).

tBMBT may provide the first experimental platform for isolated Moir\'e flat Chern bands. Besides tBMBT, there are rich choices of magnetic layered topological materials such as Mn$_2$Bi$_2$Te$_5$~\cite{zhang2019} and MnBi$_4$Te$_7$~\cite{wuj2019}, etc. These materials provide fertile playground for investigating emergent correlated topological states in twisted multilayers with tunable $U/W$.

\begin{acknowledgments}
\emph{Acknowledgments.} B.L. is supported by Princeton Center for Theoretical Science at Princeton University. Y.Z. acknowledges support from National Key Research Program of China (grant nos. 2016YFA0300703, 2018YFA0305600), NSF of China (grant nos. U1732274, 11527805, 11425415 and 11421404), and Strategic Priority Research Program of Chinese Academy of Sciences (grant no. XDB30000000). J.W. is supported by the Natural Science Foundation of China through Grant No.~11774065, the National Key Research Program of China under Grant No.~2016YFA0300703, the Natural Science Foundation of Shanghai under Grant Nos.~17ZR1442500, 19ZR1471400.
\end{acknowledgments}

\bibliography{TMBT_ref}

\begin{widetext}

\section*{Supplementary Material for "Flat Chern Band From Twisted Bilayer MnBi$_2$Te$_4$"}

\section{The continuum model of tBMBT and parameters, band structure calculation}\label{sec-para}

In this section, we give the interlayer hopping Hamiltonian of the continuum model of tBMBT, and list the parameters for FM and AFM phases we fitted from the first-principles calculations~\cite{zhangd2019}. More details of the continuum model is given in Sec. \ref{sec-CM}.

The interlayer Moir\'e hopping potential $T(\mathbf{r})$ is as shown in the main text Eq. (5). Except for the zero Fourier component $T_0$, we have kept the lowest six nonzero Fourier components of the Moir\'e potential at momenta
\begin{equation}\label{Seq-g6}
\begin{split}
&\bm{g}_1=k_m(1,0),\quad \bm{g}_2=k_m(1/2,\sqrt{3}/2),\quad \bm{g}_3=k_m(-1/2,\sqrt{3}/2),\\
&\bm{g}_4=k_m(-1,0),\quad \bm{g}_5=k_m(-1/2,-\sqrt{3}/2),\quad \bm{g}_6=k_m(1/2,-\sqrt{3}/2),
\end{split}
\end{equation}
where $k_m=8\pi\sin(\theta/2)/\sqrt{3}a_0$ is the length of the Moir\'e reciprocal vector. Higher Fourier components are expected to decay exponentially with the Fourier momentum \cite{bistritzer2011}, therefore can be ignored.

For FM phase tBMBT, the interlayer hopping matrices can be divided into a nonmagnetic part and an FM part, namely, $T_j=T_j^\text{N}+\gamma_fT_j^\text{FM}$ ($0\le j\le6$). The dimensionless parameter $\gamma_f$ is so chosen that $\gamma_f=1$ corresponds to the phase transition point of an untwisted FM bilayer MnBi$_2$Te$_4$ from trivial insulator to a Chern insulator of Chern number $1$. The zero Fourier component matrices $T_0^\text{N}$ and $T_0^\text{FM}$ have the form
\begin{equation}\label{Seq-T0FM}
T_0^\text{N}=\left(
\begin{array}{cccc}
t_1&&-i\lambda&\\
& t_2&&i\lambda\\
-i\lambda&&t_2&\\
&i\lambda&&t_1\\
\end{array}\right)\ ,\qquad
T_0^\text{FM}=\left(
\begin{array}{cccc}
u_1&&-i\kappa&\\
& -u_2&&-i\kappa\\
-i\kappa&&u_2&\\
&-i\kappa&&-u_1\\
\end{array}\right)\ ,
\end{equation}
while the nonzero Fourier component matrices $T_j^\text{N}$ and $T_j^\text{FM}$
\begin{equation}\label{Seq-TjFM}
T_j^\text{N}=\left(
\begin{array}{cccc}
\omega_jt_1'&&-i\lambda'&\\
& t_2'&&i\lambda'\\
-i\lambda'&&t_2'&\\
&i\lambda'&&\omega_jt_1'\\
\end{array}\right)\ ,\qquad
T_j^\text{FM}=\left(
\begin{array}{cccc}
\omega_ju_1'&&-i\kappa'&\\
& -u_2'&&-i\kappa'\\
-i\kappa'&&u_2'&\\
&-i\kappa'&&-\omega_ju_1'\\
\end{array}\right)\ ,\quad (1\le j\le 6)
\end{equation}
where $\omega_j=e^{i(-1)^{j}\frac{2\pi}{3}}$ is the phase factor due to the relative in-plane shift between the closest Bi atoms of the two SLs.

\begin{table}
  \centering
  \begin{tabular}{c|c|c}
  \hline
  tBMBT parameters & FM phase \ & AFM phase \\
  \hline
  $\gamma$ (eV$\cdot$\AA$^2$) & $12.8$ & $17.0$ \\
  \hline
  $m_0$ (eV) & $-0.0863$ & $-0.132$ \\
  \hline
  $\beta_0$ (eV$\cdot$\AA$^2$) & $6.72$ & $9.40$ \\
  \hline
  $\alpha$ (eV$\cdot$\AA) & $1.85$ & $3.20$ \\
  \hline
  $m_1$ (eV) & $-0.176$ & $0.05$ \\
  \hline
  $\beta_1$ (eV$\cdot$\AA$^2$) & $8.68$ &\\
  \hline
  $m_2$ (eV) & $-0.0116$ & $0.12$\\
  \hline
  $\beta_2$ (eV$\cdot$\AA$^2$) & $6.49$ & \\
  \hline
  $\alpha'$ (eV$\cdot$\AA) & $-0.532$ &\\
  \hline
  $t_1$ (eV) & $-0.0202$ & $-0.0444$ \\
  \hline
  $t_1'$ (eV) & $0.00086$ & $0.00073$ \\
  \hline
  $t_2$ (eV) & $0.0160$ & $0.0386$ \\
  \hline
  $t_2'$ (eV) & $0.00025$ & $0.0011$ \\
  \hline
  $\lambda$ (eV) & $0.0470$ & $0.0530$ \\
  \hline
  $\lambda'$ (eV) & $0.0010$ & $0.0020$ \\
  \hline
  $u_1$ (eV) & $-0.0068$ & \\
  \hline
  $u_1'$ (eV) & $-0.00002$ & \\
  \hline
  $u_2$ (eV) & $0.0053$ & \\
  \hline
  $u_2'$ (eV) & $-0.0015$ & \\
  \hline
  $\kappa$ (eV) & $0.0101$ & \\
  \hline
  $\kappa'$ (eV) & $0.00339$ & \\
  \hline
  \end{tabular}
  \caption{Parameters for the tBMBT continuum model in the FM phase and the AFM phase.}\label{Tab-tBMBT}
\end{table}

\begin{figure}[tbp]
\begin{center}
\includegraphics[width=6.5in]{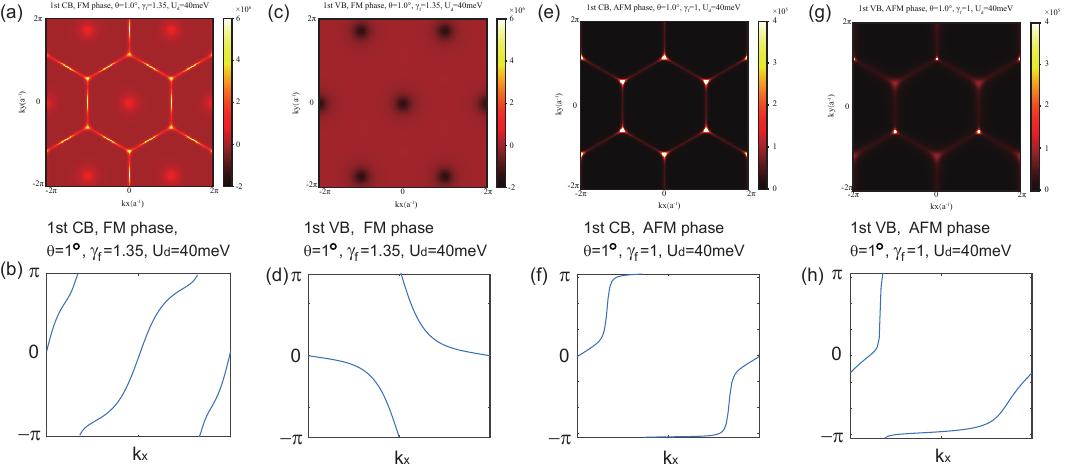}
\end{center}
\caption{The Berry curvature and Wilson loop eigenvalue for the 1st conduction band and the 1st valence band, where the parameters are (a)-(d) FM phase, $\theta=1^\circ$, $U_d=40$meV, $\gamma_f=1.02$ (band structure given in main text Fig. 2(b)), and (e)-(h) AFM phase, $\theta=1^\circ$, $\gamma_{af}=0.75$, $U_d=40$meV (band structure given in main text Fig. 3(b)).
}
\label{figSWL}
\end{figure}

In the AFM phase tBMBT, the interlayer hopping matrices $T_j$ take the form $T_j=T_j^\text{N}$ as allowed by symmetries, where $T_j^\text{N}$ are still given by Eqs. (\ref{Seq-T0FM}) and (\ref{Seq-TjFM}). The derivation of these matrices can be found in Sec. \ref{sec-CM}.

The parameters of the monolayer Hamiltonian in the main text Eqs. (3) and (4), and the interlayer hopping parameters in Eqs. (\ref{Seq-T0FM}) and (\ref{Seq-TjFM}) for the FM and the AFM phases tBMBT are listed in Tab. \ref{Tab-tBMBT}, which are estimated from the first-principles calculations (see Secs. \ref{sec-KP} and \ref{sec-CM}).

The band structure is calculated by Fourier transforming the continuum model in main text Eq. (1) into the momentum space \cite{bistritzer2011}. The transformation brings the Hamiltonian into a momentum space hopping model
\begin{equation}
H_{\mathbf{Q}\mathbf{Q}'}(\mathbf{k})=\left(\begin{array}{cc}
h_{1,\frac{\theta}{2}}(\mathbf{k}+\mathbf{Q})+U_d& T_0\\
T^\dag_0& h_{2,-\frac{\theta}{2}}(\mathbf{k}+\mathbf{Q})-U_d\\
\end{array}\right)\delta_{\mathbf{Q},\mathbf{Q}'} + \Big[\sum_{j=1}^6\left(\begin{array}{cc}
0& T_j\\
0&0\\
\end{array}\right)\delta_{\mathbf{Q},\mathbf{Q}'+\bm{g}_j}+h.c.\Big]\ ,
\end{equation}
where $\mathbf{Q}=m_1\bm{g}_1+m_2\bm{g}_2$ ($m_1,m_2\in\mathbb{Z}$) runs over all reciprocal lattice sites, and $\mathbf{k}$ is in the Moir\'e BZ. To do numerical calculations, one can take a cutoff in reciprocal lattice $\mathbf{Q}$. We note that the form of the monolayer $\mathbf{k}\cdot \mathbf{p}$ Hamiltonian $h_{l,\frac{\theta}{2}}(\mathbf{k})$ given in the main text is only valid for small $\mathbf{k}$. At large $\mathbf{k}$, the dispersions of $h_{l,\frac{\theta}{2}}(\mathbf{k})$ may severely deviate from the actual monolayer band structure and incorrectly disperse into the CNP gap, which will affect the low energy band structure. To avoid this, one can either take a small enough cutoff in $\mathbf{Q}$ so that large $\mathbf{k}$ is not involved, or correct the large $\mathbf{k}$ dispersions by adding proper higher power terms of $\mathbf{k}$ in $h_{l,\frac{\theta}{2}}(\mathbf{k})$. The low energy physics is not affected by the $\mathbf{Q}$ cutoff or large $\mathbf{k}$ dispersion corrections.

The Chern numbers of the bands can be calculated by either the Wilson loop winding number \cite{alex2014} or the integration of Berry curvature \cite{tknn1982} in the Moir\'e BZ. Figs. \ref{figSWL}  shows two examples for the band structures of main text Fig. 2(b) and Fig. 3(b), respectively, where we have plotted the Berry curvature in the Moir\'e BZ and the Wilson loop eigenvalues sweeping across the Moir\'e BZ for the 1st conduction band and 1st valence band. The Chern number is simply equal the the Wilson loop winding number. Fig. \ref{figs2} shows another example in the AFM phase at larger angle $\theta=2^\circ$, where the first valence band carries Chern number $-2$.

\begin{figure}[tbp]
\begin{center}
\includegraphics[width=6.0in]{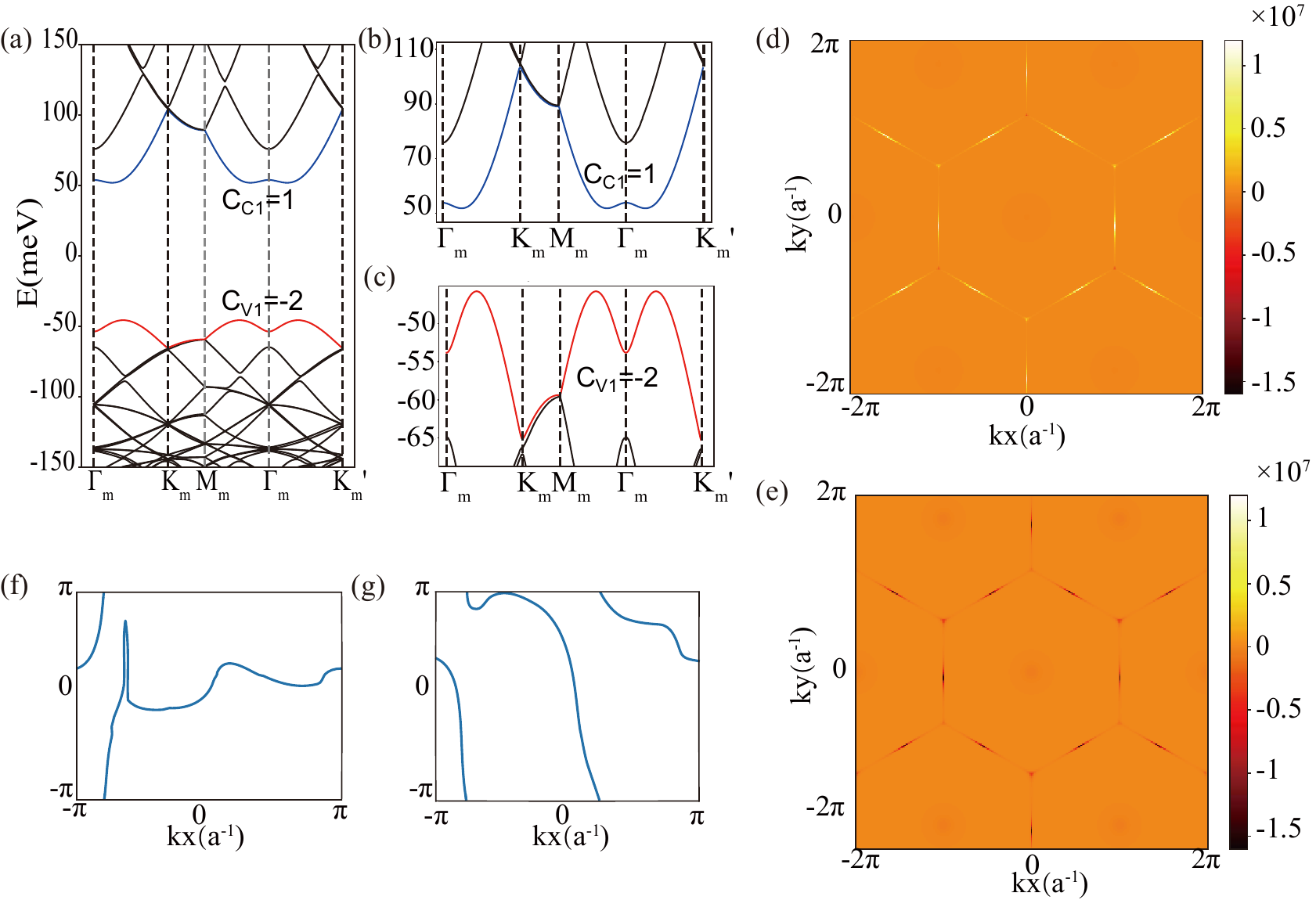}
\end{center}
\caption{The band structure, Berry curvature and Wilson loop eigenvalue for the 1st conduction band and the 1st valence band of AFM phase with the parameters $\theta=2^\circ$, $U_d=40$meV,$\gamma_{af}=1.35$. (a) The band structure. (b)-(c) Zoom in plot of (a) showing the bandwidth of the 1st conduction band and 1st valance band. (d)-(e) The Berry curvature for the 1st conduction band and the 1st valance band. (f)-(g) The Wilson loop eigenvalue for the 1st conduction band and the 1st valance band.}
\label{figs2}
\end{figure}

\section{The 3D bulk MnBi$_2$Te$_4$ Model}\label{sec-KP}

The form and parameters of our tBMBT model for two SLs of MnBi$_2$Te$_4$ is derived from the reduction of the 3D bulk MnBi$_2$Te$_4$ model into two-dimensional SLs, which we briefly review and describe in this section.

MnBi$_2$Te$_4$ has a seven layer structure (Te-Bi-Te-Mn-Te-Bi-Te), which gives a septuple layer (SL). Each atomic layer forms a triangular lattice, while each SL forms an ABCABCA stacking (from the bottom layer to the top layer). The in-plane lattice constant is $a_0=4.334$\AA, and the $z$ direction thickness of each unit cell (which consists of 3 SLs) is $c_0=40.91$\AA.

Here we present two kinds of crystal structures: (1) the ground state structure which has the SLs forming AB stacking, namely, the stacking of the atomic layers is ABCABCA-BCABCAB-CABCABC-$\cdots$. (2) the unstable structure where SLs form AA stacking, namely, atomic layer stacking ABCABCA-ABCABCA-$\cdots$.

In both AB stacking and AA stacking, the generic nonmagnetic 3D bulk MnBi$_2$Te$_4$ has an effective $\mathbf{k}\cdot \mathbf{p}$ Hamiltonian for momentum $\mathbf{k}$ measured from the $\Gamma$ point \cite{zhangd2019}:
\begin{equation}\label{Seq-HN}
H_{\text{N}}(\mathbf{k})=E_0(\mathbf{k})\mathbf{1}_{4}+\left(
\begin{array}{cccc}
M(\mathbf{k})& A_2 k_-&A_1k_z&\\
A_2k_+& -M(\mathbf{k}) &&-A_1k_z\\
A_1k_z&&-M(\mathbf{k})&A_2 k_-\\
&-A_1k_z&A_2 k_+&M(\mathbf{k})\\
\end{array}\right)\ ,
\end{equation}
where $\mathbf{1}_4$ is the $4\times 4$ identity matrix, $E_0(\mathbf{k})=C+D_1k_z^2+D_2(k_x^2+k_y^2)$, $M(\mathbf{k})=M_0+B_1k_z^2+B_2(k_x^2+k_y^2)$, and $k_\pm=k_x\pm ik_y$. The basis of the Hamiltonian is given by
\[\Psi_\Gamma=(|p_{z,\text{Bi}}^{+},\uparrow\rangle, |p_{z,\text{Te}}^{-},\downarrow\rangle, |p_{z,\text{Te}}^{-},\uparrow\rangle, |p_{z,\text{Bi}}^{+},\downarrow\rangle)^T\ ,\]
where $|p_{z,\text{Bi}}^{+},s_z\rangle$ is the bonding state of $p_z$ orbitals of the two Bi layers in a unit cell with spin $s_z=\uparrow,\downarrow$, and $|p_{z,\text{Te}}^{-},s_z\rangle$ is the anti-bonding state of $p_z$ orbitals of the top and bottom Te layers in a unit cell with spin $s_z$. The symmetries of the 3D nonmagnetic phase are the time reversal $T$, the inversion $P$, the 3-fold rotation about $z$ axis $C_{3z}$ and the 2-fold rotation about $x$ axis $C_{2x}$. Their symmetry operations are given by $TH_{\text{N}}(\mathbf{k})T^{-1}=U_TH_{\text{N}}^*(-\mathbf{k})U_T^\dag$, $PH_{\text{N}}(\mathbf{k})P^{-1}=U_PH_{\text{N}}(-\mathbf{k})U_P^\dag$, $C_{3z}H_{\text{N}}(\mathbf{k})C_{3z}^{-1}=U_{C_{3z}}H_{\text{N}}(C_{3z}\mathbf{k})U_{C_{3z}}^\dag$ and
$C_{2x}H_{\text{N}}(\mathbf{k})C_{2x}^{-1}=U_{C_{3z}}H_{\text{N}}(C_{2x}\mathbf{k})U_{C_{2x}}^\dag$, where
\begin{equation}
U_T=\left(\begin{array}{cccc}
&&&-1\\
&&1&\\
&-1&&\\
1&&&\\
\end{array}\right),\
U_P=\left(\begin{array}{cccc}
1&&&\\
&-1&&\\
&&-1&\\
&&&1\\
\end{array}\right),\
U_{C_{3z}}=\left(\begin{array}{cccc}
e^{\frac{i \pi}{3}}&&&\\
&e^{-\frac{i \pi}{3}}&&\\
&&e^{\frac{i \pi}{3}}&\\
&&&e^{-\frac{i \pi}{3}}\\
\end{array}\right),\
U_{C_{2x}}=\left(\begin{array}{cccc}
&&&i\\
&&-i&\\
&-i&&\\
i&&&\\
\end{array}\right),
\end{equation}
respectively.

The layered structure of MnBi$_2$Te$_4$ makes the bands less dispersive in the $k_z$ direction. To approximately recover the real space layered structure in the $z$ direction, we can apply the substitution
\begin{equation}\label{sub-kz}
k_z\rightarrow \frac{1}{c_0}\sin (k_z c_0)\ ,\qquad k_z^2\rightarrow \frac{2}{c_0^2}[1-\cos (k_z c_0)]
\end{equation}
in the $\mathbf{k}\cdot \mathbf{p}$ Hamiltonian $H_{\text{N}}(\mathbf{k})$. In this way, we have a Hamiltonian periodic in $k_z$, i.e., $H_{\text{N}}(\mathbf{k})=H_{\text{N}}(\mathbf{k}+\frac{2\pi}{c_0}\hat{\mathbf{z}})$, and we can extract out the Hamiltonian within each SL and the hoppings between neighbouring SLs. We shall apply this substitution in the following.

At low temperatures, either ferromagnetism (FM) or anti-ferromagnetism (AFM) is developed in MnBi$_2$Te$_4$. In the below, we describe their effective models, respectively.


(1) {\bf The FM phase.} In this phase, the system develops a uniform FM order in the $z$ direction. When the FM order is in the $+z$ direction, the 3D $\mathbf{k}\cdot \mathbf{p}$ Hamiltonian at $\Gamma$ point becomes $H_{\text{FM}}(\mathbf{k})=H_{\text{N}}(\mathbf{k})+H_{\text{FM}}(\mathbf{k})$, where
\begin{equation}
H_{\text{FM}}(\mathbf{k})=\left(
\begin{array}{cccc}
M_{1}(\mathbf{k})& A_4 k_-&A_3k_z&\\
A_4k_+& -M_2(\mathbf{k}) &&A_3k_z\\
A_3k_z&&M_2(\mathbf{k})&-A_4 k_-\\
&A_3k_z&-A_4 k_+&-M_1(\mathbf{k})\\
\end{array}\right)\ .
\end{equation}
Here we have defined $M_1(\mathbf{k})=M_1+B_3k_z^2+B_4(k_x^2+k_y^2)$, and $M_2(\mathbf{k})=M_2+B_5k_z^2+B_6(k_x^2+k_y^2)$. If the FM order is in the $-z$ direction, $H_{\text{FM}}$ flips its sign. Similarly, we apply the substitution (\ref{sub-kz}) to obtain the layered Hamiltonian of the FM phase.

(2) {\bf The AFM phase.} In this case, the system develops a FM order in the $\pm z$ direction in each SL, while two neighbouring SLs have opposite FM orders. Accordingly, the unit cell is doubled in the $z$ direction, and the Brillouin zone (BZ) is reduced by $1/2$ in the $k_z$ direction.

The AFM order yields an AFM term in the $l_z$-th SL
to be in the following form:
\begin{equation}
H_{\text{AFM}}(\mathbf{k}_\parallel,l_z)=(-1)^{l_z}\left(
\begin{array}{cccc}
M_{1}&&&\\
& -M_2 &&\\
&&M_2&\\
&&&-M_1\\
\end{array}\right)\ ,
\end{equation}
where $M_1$ and $M_2$ are exchange fields, and we have ignored the $\mathbf{k}$ dependence of the AFM term (which is difficult to obtain accurately from first-principles). The full Hamiltonian of the AFM phase is given by $H_{\text{AFM}}=H_{\text{N}}+H_{\text{AFM}}$. The AFM term $H_{\text{AFM}}$ induces a hopping between 3D momenta $\mathbf{k}$ and $\mathbf{k}+\frac{\pi}{c_0}\hat{\mathbf{z}}$. This opens two gaps of approximate magnitudes $2M_1$ and $2M_2$ at $k_z=\pm \frac{\pi}{2c_0}$, and reduces the BZ size by one half in the $k_z$ direction.


The parameters for the AA stacking and AB stacking 3D MnBi$_2$Te$_4$ in the FM and AFM phases can be defetermined from first-principles calculations~\cite{zhangd2019}.

\section{The continuum model of twisted bilayer MnBi$_2$Te$_4$}\label{sec-CM}

In this section, we obtain the single particle continuum model for the twisted bilayer MnBi$_2$Te$_4$ (tBMBT) Moir\'e pattern from the $\mathbf{k}\cdot \mathbf{p}$ Hamiltonian of the 3D MnBi$_2$Te$_4$ in Sec. \ref{sec-KP}. Hereafter we use $\mathbf{k}=(k_x,k_y)$ to denote the 2D momentum.


Using substitution (\ref{sub-kz}), we can separate the 3D $\mathbf{k}\cdot \mathbf{p}$ Hamiltonian into the monolayer Hamiltonian in each SL and the $z$ direction hopping terms between neighboring SLs. The monolayer Hamiltonian of the $l$'s layer ($l=1,2$) takes the form $h_l(\mathbf{k})=h_\text{N}(\mathbf{k})+\gamma_f h_{\text{FM}}(\mathbf{k})$ for FM state, and $h_l(\mathbf{k})=h_\text{N}(\mathbf{k})-\gamma_{af}(-1)^l h_{\text{AFM}}(\mathbf{k})$, where $\gamma_f$ and $\gamma_{af}$ are two dimensionless parameters for tuning the strength of the FM and AFM orders, respectively, while $h_{\text{FM}}(\mathbf{k})$ and $h_{\text{AFM}}(\mathbf{k})$ are fixed. Since the definitions of $\gamma_f$ and $\gamma_{af}$ are up to a rescaling, we need to fix a reference point for them. For the FM phase, we fix $\gamma_f=1$ to be the critical point for an untwisted bilayer MnBi$_2$Te$_4$ to undergo the transition from a trivial insulator (where $\gamma_f<1$) to a Chern insulator of Chern number $1$ (where $\gamma_f>1$). With this choice of reference point, we find the FM order strength given by our first-principles calculations for 3D bulk MnBi$_2$Te$_4$ corresponds to $\gamma_f=\gamma_{f0}=0.75$. The realistic FM order strength may differ from the value given by first-principles calculations, and is tunable by a small external magnetic field. For the AFM phase, we simply fix $\gamma_{af}=1$ to be the AFM strength obtained from our first-principles calculations.

The normal term is given by
\begin{equation}
h_\text{N}(\mathbf{k})=\epsilon_0(\mathbf{k})\mathbf{1}_{4}+\left(
\begin{array}{cccc}
m(\mathbf{k})& \alpha k_-&&\\
\alpha k_+& -m(\mathbf{k}) &&\\
&&-m(\mathbf{k})&\alpha k_-\\
&&\alpha k_+&m(\mathbf{k})\\
\end{array}\right)\ ,
\end{equation}
where $\epsilon_0(\mathbf{k})=\epsilon_0+\gamma \mathbf{k}^2$, and $m(\mathbf{k})=m_0+\beta_0 \mathbf{k}^2$. The parameters obtained in this way are related to the 3D bulk parameters in Sec. \ref{sec-KP} (for a particular stacking structure) by $\epsilon_0=C+\frac{2D_1}{c_0^2}$, $\gamma=D_2$, $m_0=M_0+\frac{2B_1}{c_0^2}$, $\beta_0=B_2$, and $\alpha=A_2$. Since $\epsilon_0$ is a constant term, we can set $\epsilon_0=0$ by redefining the zero point of energy, so we have $\epsilon_0(\mathbf{k})=\gamma \mathbf{k}^2$. 

The FM term is given by
\begin{equation}
h_{\text{FM}}(\mathbf{k})=\left(
\begin{array}{cccc}
m_{1}(\mathbf{k})& \alpha' k_-&&\\
\alpha'k_+& -m_2(\mathbf{k}) &&\\
&&m_2(\mathbf{k})&-\alpha' k_-\\
&&-\alpha' k_+&-m_1(\mathbf{k})\\
\end{array}\right)\ ,
\end{equation}
where $m_1(\mathbf{k})=m_1+\beta_1 \mathbf{k}^2$, and $m_2(\mathbf{k})=m_2+\beta_2 \mathbf{k}^2$. The parameters are approximately $m_1=\gamma_{f0}^{-1}(M_1+\frac{2B_3}{c_0^2})$, $m_2=\gamma_{f0}^{-1}(M_2+\frac{2B_5}{c_0^2})$, $\beta_1=\gamma_{f0}^{-1}B_4$, $\beta_2=\gamma_{f0}^{-1}B_6$, and $\alpha'=\gamma_{f0}^{-1}A_4$, where $\gamma_{f0}=0.75$ is the dimensionless parameter characterizing the FM order strength of the 3D first-principles result we defined earlier.

The AFM term is given by
\begin{equation}
h_{\text{AFM}}(\mathbf{k})=\left(
\begin{array}{cccc}
m_{1}&&&\\
& -m_2&&\\
&&m_2&\\
&&&-m_1\\
\end{array}\right)\ ,
\end{equation}
where $m_1$ and $m_2$ are constants. In this case, one has $m_1=M_1$, and $m_2=M_2$ (since $\gamma_{af}=1$ is fixed to be the AFM order strength given by our 3D first-principles calculations). Besides, one could find out the interlayer hopping matrix for both the AA stacking and the AB stacking configuration.

In the tBMBT, there are both AA stacking positions and AB/BA stacking positions. As a commonly used approximation, we can assume the monolayer Hamiltonian is only a function of momentum $-i\nabla$ and is independent of positions, but is rotated by angle $\pm\theta/2$ in the first and second layers, respectively. Namely, $h_{l,\pm\theta/2}(\mathbf{k})=R_{\pm\frac{\theta}{2}}^\dag h_l(\mathbf{k})R_{\pm\frac{\theta}{2}}$, where $R_{\pm\frac{\theta}{2}}=\text{diag}(e^{\pm i\theta/4},e^{\mp i\theta/4},e^{\mp i\theta/4},e^{\pm i\theta/4})$ is the rotation matrix. We estimate the parameters of the tBMBT monolayer Hamiltonian by properly averaging between the AB stacking and AA stacking parameters. 


Meanwhile, the interlayer hopping is a function of position $\mathbf{r}$, with a spatial period given by the Moire superlattice. Keeping the lowest six nonzero Fourier components, we can write the interlayer Moir\'e hopping potential $T(\mathbf{r})$ from layer $2$ to layer $1$ in the form
\begin{equation}
T(\mathbf{r})=T_0+\sum_{j=1}^6 T_j e^{i\bm{g}_j\cdot\mathbf{r}}\ ,
\end{equation}
where $T_0$ and $T_j$ ($1\le j\le 6$) are $4\times 4$ hopping matrices, and $\bm{g}_j$ ($1\le j\le 6$) are the six smallest Moir\'e reciprocal vectors in Eq. (\ref{Seq-g6}). We assume $\mathbf{r}=0$ corresponds to AA stacking. In this convention, the AB stacking positions $\mathbf{r}_{AB}$ satisfy $e^{i\bm{g}_j\cdot\mathbf{r}_{AB}}=e^{-i(-1)^{j}\frac{2\pi}{3}}$. From the substitution (\ref{sub-kz}), we find the hopping matrices for the FM phase with FM order strength $\gamma_f$ are of the form
\begin{equation}\label{seq-T0Tj}
\begin{split}
&T_0=\left(
\begin{array}{cccc}
t_1+\gamma_fu_1&0&-i(\lambda+\gamma_f\kappa)&0\\
0& t_2-\gamma_fu_2&0&i(\lambda-\gamma_f\kappa)\\
-i(\lambda+\gamma_f\kappa)&0&t_2+\gamma_fu_2&0\\
0&i(\lambda-\gamma_f\kappa)&0&t_1-\gamma_fu_1\\
\end{array}\right),\\
\ \\
&T_j=\left(
\begin{array}{cccc}
(t_1'+\gamma_fu_1')\omega_j&0&-i(\lambda'+\gamma_f\kappa')&0\\
0& t_2'-\gamma_fu_2'&0&i(\lambda'-\gamma_f\kappa')\\
-i(\lambda'+\gamma_f\kappa')&0&t_2'+\gamma_fu_2'&0\\
0&i(\lambda'-\gamma_f\kappa')&0&(t_1'-\gamma_fu_1')\omega_j\\
\end{array}\right),
\end{split}
\end{equation}
where $\omega_j=e^{i(-1)^{j}\frac{2\pi}{3}}$. They can then be decomposed into the nonmagnetic part and FM part (if FM phase) as shown in Eqs. (\ref{Seq-T0FM}) and (\ref{Seq-TjFM}). For the AFM phase, the hopping matrices are given by setting $u_j=u_j'=\kappa=\kappa'=0$ in Eq. (\ref{seq-T0Tj}). The factor $\omega_j$ is the phase factor due to the relative in-plane shift between closest Bi atoms in neighboring SLs (assume the hopping is dominantly between the closest atoms). This is because at AA stackings, the closest Bi atoms of the orbital $|p_{z,\text{Bi}}^{+},s\rangle$ in the two SLs are not on top of each other but shifted in-plane by $1/3$ unit cell, while at AB stackings, the closest Bi atoms in the two SLs are on top of each other. In contrast, the closest Te atoms in AA stacking are on top of each other, while shifted in-plane by $1/3$ unit cell at AB stacking.

The coefficient can be determined by let $T(\mathbf{r})$ at AA stacking ($\mathbf{r}=0$) and AB stacking ($e^{i\bm{g}_j\cdot\mathbf{r}}=\omega_j^{-1}$) positions agree with that of untwisted AA stacking and untwisted AB stacking structures, respectively (which is a good approximation for small twist angle \cite{bistritzer2011}). For the FM phase, they are roughly related to the 3D parameters by
\begin{equation}\label{seq-para}
\begin{split}
&t_1-3t_1'=-\frac{2(B_1+D_1)|_{AA}}{c_0^2}\ ,\quad t_1+6t_1'=-\frac{2(B_1+D_1)|_{AB}}{c_0^2}\ ,\quad u_1-3u_1'=-\frac{2B_3|_{AA}}{\gamma_{f0}c_0^2}\ ,\quad u_1+6u_1'=-\frac{2B_3|_{AB}}{\gamma_{f0}c_0^2}\ ,\\
&t_2+6t_2'=-\frac{2(D_1-B_1)|_{AA}}{c_0^2}\ ,\quad t_2-3t_2'=-\frac{2(D_1-B_1)|_{AB}}{c_0^2}\ ,\quad u_2+6u_2'=-\frac{2B_5|_{AA}}{\gamma_{f0}c_0^2}\ ,\quad u_2-3u_2'=-\frac{2B_5|_{AB}}{\gamma_{f0}c_0^2}\ ,\\
&\lambda+6\lambda'=\frac{A_1|_{AA}}{c_0}\ ,\quad \lambda-3\lambda'=\frac{A_1|_{AB}}{c_0}\ ,\quad \kappa+6\kappa'=\frac{A_3|_{AA}}{\gamma_{f0}c_0}\ ,\quad \kappa-3\kappa'=\frac{A_3|_{AB}}{\gamma_{f0}c_0}\ ,\\
\end{split}
\end{equation}
where $\gamma_{f0}=0.75$ is the FM strength corresponding to the 3D bulk first-principles result. For the AFM phase, $u_j=u_j'=\kappa=\kappa'=0$, while the other parameters are given by Eq. (\ref{seq-para}). The continuum model Hamiltonian of tBMBT can then be written as
\begin{equation}
H=\left(
\begin{array}{cc}
h_{1,\theta/2}(-i\nabla)& T(\mathbf{r})\\
T^\dag(\mathbf{r})& h_{2,-\theta/2}(-i\nabla)\\
\end{array}
\right)\ ,
\end{equation}
as we have in the main text.

\section{Discussion on the FM and AFM phases}
Since the 3D MnBi$_2$Te$_4$ is in the AFM phase, it is more likely that the tBMBT ground state is AFM as well. In this case, it is expected that the AFM phase can be easily flipped into the FM phase by a magnetic field. In first principles calculations, the AFM and the FM phases of MnBi$_2$Te$_4$ have competing energies. 

Experimentally, the AFM phase of MnBi$_2$Te$_4$ is shown to undergo a spin-flop transition into the FM phase around $3$T \cite{otrokov2018,deng2019}. Theoretically, the AFM exchange interaction between neighboring SLs is estimated by ab initio calculations to be $J_\perp\approx -0.02$ meV$/\mu_B^2$ \cite{otrokov2018}, while the magnetic moment of Mn atom is $5\mu_B$. Therefore, the magnetic field for polarizing the system into the FM phase can be estimated to be $B_{c}\approx |J_\perp|\times 5\mu_B\approx 2$T. Therefore, we conclude that the AFM phase of tBMBT can be easily flipped into the FM phase by a magnetic field around $2\sim4$T.

In addition, it might also be possible to control the magnetization of tBMBT by adding top and bottom FM heterostructures, which may induce proximity exchange couplings. We leave the investigation of this possibility to the future studies.

\end{widetext}

\end{document}